\documentclass[aps,prl,twocolumn,showpacs,superscriptaddress, longbibliography]{revtex4-1}

\usepackage{xcolor, graphicx, color, placeins}
\usepackage{amssymb,amsmath, amsfonts, bm}
\usepackage{newtxtext,newtxmath}
\usepackage{dcolumn}
\usepackage[none]{hyphenat} 
\usepackage{tabularx}
\usepackage{titletoc}

\usepackage{hyperref}
\hypersetup{
    colorlinks=true,
    linkcolor=blue,    
    urlcolor=blue,
    citecolor=blue,
    }
    
\usepackage{multibib}
\newcites{SI}{Reference}
\usepackage{easyReview}

\begin{document}
\title{Hole distribution and self-doping enhanced electronic correlation in hole-doped infinite-layer nickelates}

\author{Hongbin Qu}
 \affiliation{School of Physical Science and Technology, ShanghaiTech University, Shanghai 201210, China}

\author{Guang-Ming Zhang}
\email{zhangguangming@shanghaitech.edu.cn}
\affiliation{School of Physical Science and Technology, ShanghaiTech University, Shanghai 201210, China}
\affiliation{State Key Laboratory of Quantum Functional Materials, ShanghaiTech University, Shanghai 201210, China}

\author{Gang Li}
 \email{ligang@shanghaitech.edu.cn}
\affiliation{School of Physical Science and Technology, ShanghaiTech University, Shanghai 201210, China}
\affiliation{\mbox{ShanghaiTech Laboratory for Topological Physics, ShanghaiTech University, Shanghai 201210, China}}
\affiliation{State Key Laboratory of Quantum Functional Materials, ShanghaiTech University, Shanghai 201210, China}

\date{\today}

\begin{abstract}
The minimal model for infinite-layer nickelates remains under debate, particularly regarding the hybridization between itinerant interstitial-$s$ and the correlated Ni-3$d_{x^2-y^2}$ orbitals, as well as the interaction between $d_{x^2-y^2}$ and other $3d$ orbitals. Additionally, how the doped holes in La$_{1-x}$Sr$_x$NiO$_2$ are distributed among different orbitals remain unresolved. Motivated by recent angle resolved photoemission spectroscopy (ARPES) experiments, we theoretically study the electronic structure of infinite-layer La$_{1-x}$Sr$_x$NiO$_2$ at various doping levels. We find that, unlike the expectation from a rigid band shift, holes are equally  distributed to Ni-3$d_{x^2-y^2}$ and interstitial-$s$ orbitals. The role of interstitial-$s$ orbital is further confirmed from the renormalization of Ni-3$d_{x^2-y^2}$ band, for which the coupling between interstitial-$s$ and Ni-3$d_{x^2-y^2}$ exerts a non-negligible impact on the orbital-selective renormalization observed in ARPES. 
We also discuss the implication of our results to the single-band model, where the interstitial-$s$ orbital in the normal state of La$_{1-x}$Sr$_x$NiO$_2$ acts as charge donator enhancing the correlation of Ni-3$d_{x^2-y^2}$ by increasing its concentration close to half-filling.
\end{abstract}

\maketitle

\textit{Introduction} --
The discovery of high-temperature superconductivity in cuprates~\cite{HTS_1986}  significantly exceeds the upper bound of superconducting transition temperature ($T_{c}$) promised by Bardeen-Cooper-Schrieffer (BCS) theory~\cite{PhysRev.108.1175}. 
It not only offers a renewed prospect for achieving room-temperature superconductivity~\cite{lechermann_late_2020, botana_similarities_2020, ding_cuprate-like_2024} but also spurs extensive efforts to identify analogous families of unconventional superconductors.
The isostructural nickelates were later proposed as a promising candidate to cuprates, which share a great similiarity on their $d^{9-x}$ valence configurations~\cite{PhysRevB.59.7901}, the partially-occupied $d_{x^2-y^2}$ orbital~\cite{jiang_electronic_2019, sakakibara_model_2020, nomura_formation_2019, adhikary_orbital-selective_2020, hepting_electronic_2020}, and the hole doping-dependent superconducting domes~\cite{PhysRevLett.125.147003, PhysRevLett.125.027001, doi:10.1126/sciadv.abl9927,10.3389/fphy.2022.834658, osada_nickelate_2021,Kyuho-Huang}, etc. 
The infinite-layer nickelates RNiO$_2$ (R = Nd, Pr, La)~\cite{li_superconductivity_2019, osada_phase_2020, https://doi.org/10.1002/adma.202104083, osada_nickelate_2021, PhysRevLett.126.197001, doi:10.1126/science.abd7726, DW_Rossi_2022, Tam:2022aa, Ding:2023aa, Gu:2020ab, Zeng:2022aa, PhysRevLett.126.087001, Lee:2023ab, PhysRevLett.129.027002}, providing another material platform mimicking many important features of cuprates holding the potential to uncover the mechanism for high-$T_{c}$ superconductivity. 
While, important differences to the cuprates also have been noted~\cite{lechermann_late_2020,  lechermann_multiorbital_2020, wang_hunds_2020, kang_optical_2021, botana_low_2022, kang_infinite-layer_2023, liao_absence_2023}. The parent state of infinite-layer nickletes is metallic irrespective of the strength of Coulomb interactions on $d_{x^2-y^2}$ orbital, due to the electron pockets formed by R-$5d$ bands~\cite{hepting_electronic_2020, lechermann_late_2020, doi:10.1073/pnas.2007683118, sakakibara_model_2020, botana_similarities_2020, leonov_lifshitz_2020, kitatani_nickelate_2020, 10.3389/fphy.2021.810394}.
The additional Fermi surface harbors nickelates potentially beyond-single-band scenario, posing challenges to understanding its electronic structure and superconducting mechanism.
Furthermore, the oxygen states in nickelates reside much deeper below the Fermi level, rendering a reduced hybridization with the Ni-$d$ states~\cite{nomura_formation_2019, jiang_electronic_2019, botana_similarities_2020, lee_infinite-layer_2004, botana_similarities_2020}. 
The Zhang-Rice singlets in cuprates~\cite{zhang_effective_1988, dagotto_correlated_1994}, which stems from the small charge transfer gap and strong hybridization between Cu-3$d_{x^2-y^2}$ and O-$p$ states below the Fermi level, is significantly suppressed in nickelates.
This may lead to a starkly different pairing mechanism in nickelate supercondutors.
Notably, the $d_{z^2}$ orbital also heavily involves in the low-energy excitations in nickelates, additionally providing multiorbital contributions to the dominating $d_{x^2-y^2}$ states. 
Both the similarity and differences highlight the importance of nickelates as an independent material family hosting the key to the understanding of the still mysterious high-$T_{c}$ superconductivity.   

Establishing the minimal model for infinite-layer nickelate superconductors has stimulated extensive debate~\cite{nomura_formation_2019, PhysRevResearch.1.032046, kitatani_nickelate_2020, PhysRevLett.124.207004,  botana_similarities_2020, lechermann_multiorbital_2020, PhysRevX.10.021061, PhysRevB.101.041104, wang_hunds_2020, zhang_self-doped_2020, adhikary_orbital-selective_2020, wu_robust_2020, kang_optical_2021, Nomura_2022, kang_infinite-layer_2023, Chen:2023aa, 10.3389/fphy.2022.834682, chen_electronic_2024, kugler_low-energy_2024, worm_spin_2024}, especially regarding the role of the so-called ``interstitial-$s$" orbital, Ni-$d_{z^2}$, and O-$p$ orbitals. 
Recent doping-control and angle-resolved photoemission spectroscopy (ARPES) measurements have confirmed the presence of interstitial-$s$ orbital~\cite{li2025observationelectridelikesstates}
and revealed a band-selective renormalization in La$_{0.8}$Sr$_{0.2}$NiO$_{2}$ and  La$_{0.8}$Ca$_{0.2}$NiO$_{2}$~\cite{ding_cuprate-like_2024, sun_electronic_2025}. 
The Ni-$d_{x^2-y^2}$ bands show a clear mass enhancement, while the interstitial-$s$ orbitals remain identical to the density-functional theory (DFT) predication. 
Tracing the distribution of the doped-holes in these orbitals and understanding their implication to the electronic structure, thus, provide a new knob for constructing the minimal model and understanding the role of interstitial-$s$ orbital.  

\begin{figure*}[t]
\centering
\includegraphics[width=\linewidth]{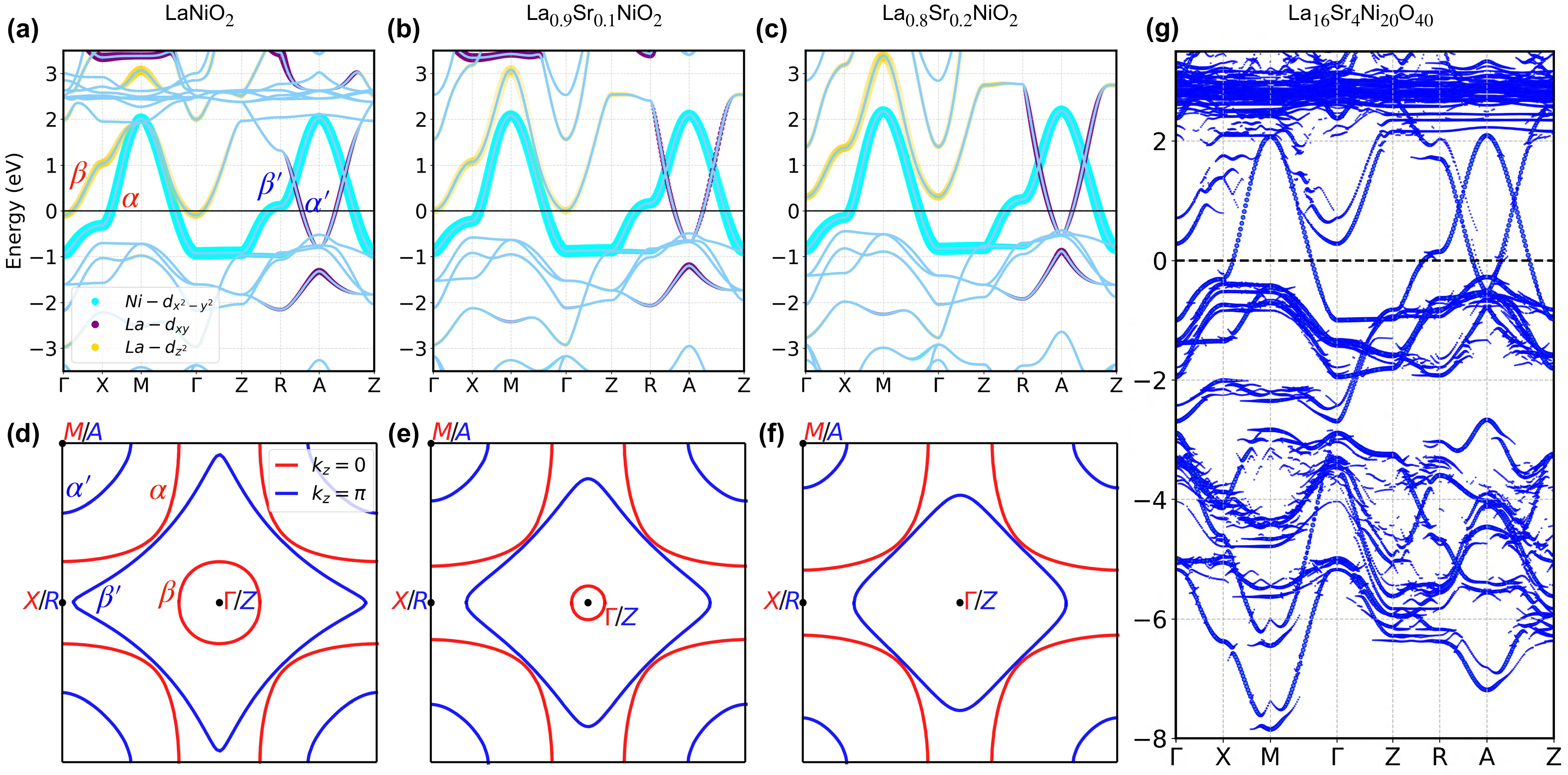}
\caption{\textbf{Electronic structure and Fermi surface of  La$_{1-x}$Sr$_{x}$NiO$_{2}$ at $\mathbf{x=0, 0.1}$, and 0.2.} (a - c) The electronic structure and its projection to Ni-$d_{x^2-y^2}$, La-$d_{xy}$, La-$d_{z^2}$ orbitals. (d - f) The corresponding Fermi surface at $k_{z} = 0$ (red line) and $k_z=\pi$ (blue line). (g) Electronic structure of La$_{0.8}$Sr$_{0.2}$NiO$_{2}$ from a slab calculation after backfolding to BZ of primitive cell.}
\label{Fig:DFT_band}
\end{figure*}

In this work, we study the hole distribution and the electronic structure of La$_{1-x}$Sr$_{x}$NiO$_{2}$ by combing DFT with dynamical mean-field theory (DMFT)~\cite{RevModPhys.68.13}. 
With the help of virtual crystal approximation (VCA)~\cite{https://doi.org/10.1002/andp.19314010507, PhysRevB.61.7877}, we evaluate the orbital occupancy of La$_{1-x}$Sr$_{x}$NiO$_{2}$ at two doping levels, i.e. $x=0.1$ and $0.2$.  
We found that the holes are mainly doped to Ni-$d_{x^2-y^2}$ and interstitial-$s$ orbitals, while the Ni-$d_{z^2}$ and O-$p$ orbitals are less affected.   
To decide the minimal model, we compare the band renormalization in two models. 
One is a six-band model with Ni 3$d$ + interstitial-$s$ orbitals. The other one is a two-band model with Ni $d_{x^2-y^2}$ + interstitial-$s$ model. 
We found that both models successfully capture the band selective renormalization consistent with the observation in ARPES. 
The hybridization and charge transfer from other Ni-$d$ orbitals contribute much less to the low-energy Ni-$d_{x^2-y^2}$ and interstitial-$s$ bands, which are, thus, likely to be less relevant to the superconductivity of infinite-layer nickelates.   
We show that further downfolding to a single-band model is possible. 
The presence of the interstitial-$s$ orbitals is found to act as a charge donator which effectively changes the doping level of the Ni-$d_{x^2-y^2}$ close to half-filling. 
Thus, counterintuitively, the presence of the metallic interstitial-$s$ orbital enhances the effective interactions between Ni-$d_{x^2-y^2}$ electrons. 
Our results highlight the unique role of the interstitial-$s$ orbitals in the construction of effective model for nickelate superconductors.  

\textit{Results}--
Figure~\ref{Fig:DFT_band} shows the Fermi surface and electronic structure of La$_{1-x}$Sr$_{x}$NiO$_{2}$ at three different doping levels. 
The VCA calculation is verified against a slab calculation with effectively $20\%$ hole doping. 
The agreement between Fig.~\ref{Fig:DFT_band}(c) and (g) confirms the reliability of VCA for this system. 
Overall, we find the electronic structure at finite hole doping follows the undoped one with the Ni-$d_{x^2-y^2}$ band crossing the Fermi level.  
With the increase of Sr concentration, the most notable change is the $\beta$ electron-pocket, which gradually shrinks and vanishes at $x > 0.1$.  
The $\beta$ pocket is mainly derived from the La-$d_{z^2}$ orbital. Some of the doped holes transfer to this orbital, removing it from the Fermi level. 
The $\alpha$ electron-pocket also shits up in binding energy leading to a closer vicinity of the van Hove singularity at $X$ to the Fermi level.
In contrast, the van Hove singularity at $R$ shifts away from the Fermi level.
The $\beta^\prime$ electron pocket also becomes smaller.  

\begin{figure*}[t]
\centering
\includegraphics[width=\linewidth]{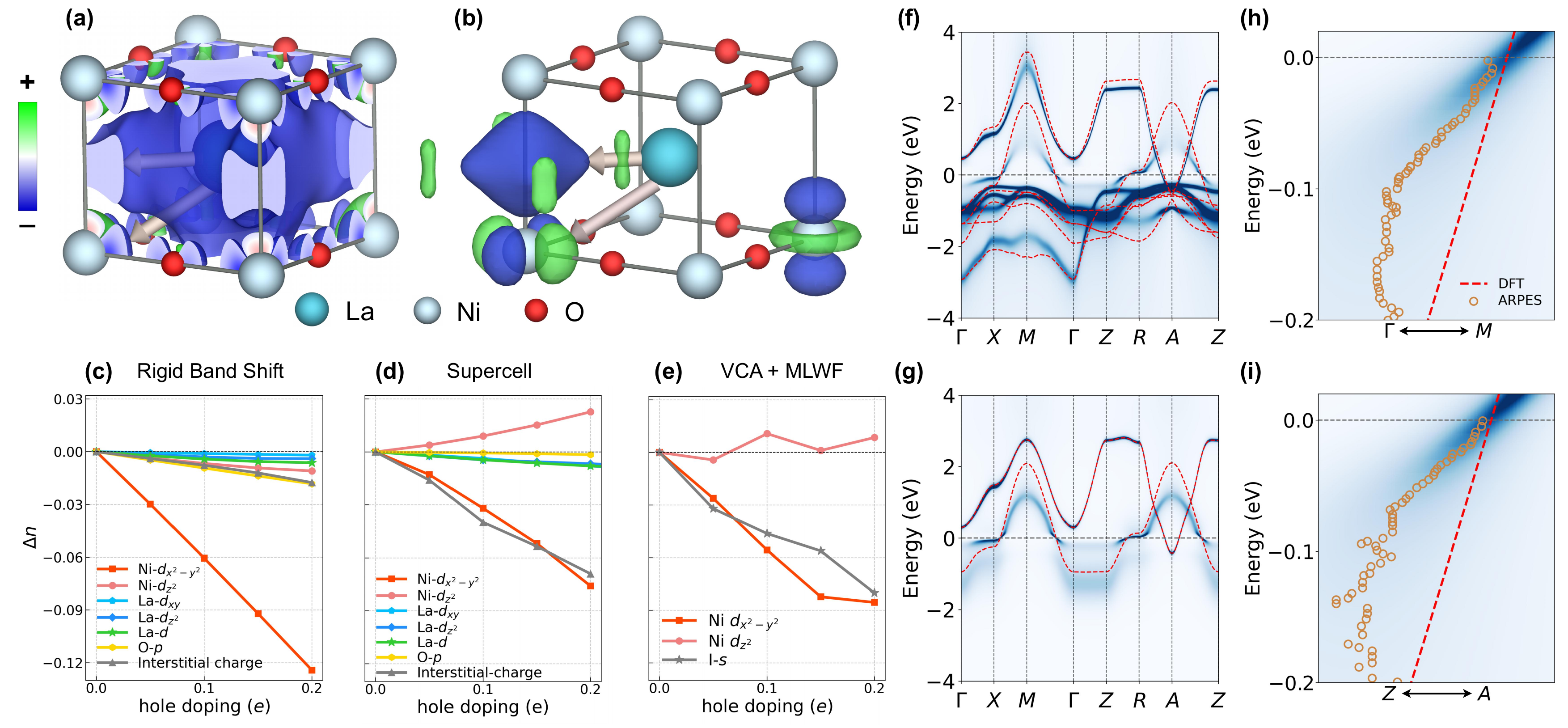}
\caption{\textbf{Hole distribution and correlated spectral function of La$_{1-x}$Sr$_{x}$NiO$_{2}$ at $x=0.2$. } (a) The differential charge density (DCD) with respect to the undoped LaNiO$_{2}$. The isocharge surface colored in blue corresponds to the holes (or missing charge) distributing in the cell. The two white thin arrows indicate that holes transfer from La to the Ni-Ni bond center, Ni, and O sites. (b) Real-space distribution of wannier orbitals obtained from a 6-band model with interstitial-$s$ and Ni-$d$ orbitals. The interstital-$s$, Ni-$d_{x^2}-y^2$, Ni-$d_{z^2}$ are displayed. (c - e) Change density change estimated from (c) a rigid-band shift in VCA primitive cell, (d) a $2\times 5\times 2$ supercell with Sr doping, and (e) the 6-band model. 
(f, g) The correlated spectral function of the (f) 6-band and (g) 2-band model, respectively. The corresponding comparison to AREPS is shown in (h) and (i). }
\label{Fig:orbital_occupancy}
\end{figure*}

Hole doping leads to a continuous evolution of the Fermi surface topology, as evident in Fig.~\ref{Fig:DFT_band}(d-f). 
The $\beta$ sheet at $k_z=0$ plane quickly vanishes as Sr concentration increases. 
In contrast, the $\alpha$ sheet around $M$ expands slowly towards $\Gamma$. 
At $k_{z}=\pi$ plane, both $\beta^\prime$ sheet around $Z$ and $\alpha^\prime$ sheet around $A$ become smaller with hole doping.  
The change of Fermi surface topology is expected to influence the electron pairing for superconductivity. 

After understanding the overall electronic structure under hole doping, we start to trace hole distribution in different orbitals.
One critical question concerning hole-doped superconductors is  where these holes are doped to. 
To answer this question, we monitor the change of electron occupancy in Ni-3$d_{x^2-y^2}$, Ni-3$d_{z^2}$, La-5$d_{xy}$, La-5$d_{z^2}$, and O-$p$ orbitals at different hole doping levels. 
The first four orbitals compose the low energy bands of  La$_{1-x}$Sr$_{x}$NiO$_{2}$. 
O-$p$ has played critical roles in cuprates but its role in nickelates is not clear. 
Figure~\ref{Fig:orbital_occupancy}(a) displays the differential charge density (DCD) between La$_{0.8}$Sr$_{0.2}$NiO$_{2}$ and undoped LaNiO$_{2}$. 
The blue isosurface corresponds to the density of the missing charge after hole-doping. 
In VCA, as we replaced La partially with Sr, the missing charge (or doped holes) should be around La before they diffuse in the system. 
While the majority of holes is around La $= (1/2, 1/2, 1/2)$, we find many holes move to  the middle of Ni-Ni bond forming the interstitial-$s$ orbital (see Fig.~\ref{Fig:orbital_occupancy}(b) for its wannier function.).   
Meanwhile, we also observe holes around Ni sites with the density isosurface displaying $d_{x^2-y^2}$ and $d_{z^2}$ atomic symmetry. 
Similarly, some holes also transfer to O site with $p$-orbital shape charge density. 
As O-$p$ mainly reside below the Fermi level, this indicates a strong Ni-$d$ and O-$p$ hybridization in the syetem, through which these holes transfer to O-$p$ orbital. 

While DCD demonstrates that doped holes mainly transfer to interstitial-$s$ and Ni-$d_{x^2-y^2}$ orbitals, it is not clear whether they are mainly doped to Ni-$d_{x^2-y^2}$ orbital as in cuprates and what the role of interstitial-$s$ orbital is.  
To answer these two questions, we design three schemes of calculation from which we can trace quantitatively the hole distribution and characterize the role of interstitial-$s$ orbital. 
In the first scheme, we rigidly shift the chemical potential down in energy to mimic hole doping. 
As evident from the electronic structure shown in Fig.~\ref{Fig:DFT_band}, rigidly shifting chemical potential down effectively dopes holes mainly to Ni-$d_{x^2-y^2}$ and La-$d_{xy}$, La-$d_{z^2}$ (interstitial-$s$) orbitals. 
We calculate the change of charge density from the orbital-projected density of states (DOS, see Supplementary Information for more details.). 
As shown in Fig.~\ref{Fig:orbital_occupancy}(c), rigidly shifting chemical potential mainly results in hole doping in Ni-$d_{x^2-y^2}$ orbital, while the charge density of other orbitals only reduces slightly.   
As the orbital-projected DOS is evaluated only inside a Wigner-Seitz (WS) cell around the given atom in DFT, the DOS of the interstitial-$s$ orbital cannot be fully accounted for in this calcualtion. 
Instead, we estimate charge density at the interstitial-$s$ orbital from the missing charge in the total density calculated in WS cell and show it as gray line with up-triangle.
As seen in Fig.~\ref{Fig:orbital_occupancy}(c), the number of holes transferring to the interstitial-$s$ orbital is much less than to the Ni-$d_{x^2-y^2}$ orbital in rigid band shift scenario. 

In the second scheme, we performed a $2\times 5 \times 2$ supercell calculation and replace one La atom with Sr. 
The corresponding charge density change is shown in Fig.~\ref{Fig:orbital_occupancy}(d). 
In contrast to the expectation from rigid band shift shown in Fig.~\ref{Fig:orbital_occupancy}(c), the doped holes nearly equally distribute to interstitial-$s$ and Ni-$d_{x^2-y^2}$ orbitals. 
Meanwhile, an effective electron doping to the Ni-$d_{z^2}$ orbital is observed indicating a charge transfer from Ni-$d_{x^2-y^2}$ to Ni-$d_{z^2}$ occurs. 

In the third scheme, we estimate the same quantity in VCA by downfolding its electronic structure to a 6-band model with Ni-3$d$ + interstitial-$s$ orbitals. 
As shown in Fig.~\ref{Fig:orbital_occupancy}(e), the main feature of the hole doping observed in supercell calculations is captured by this model, including the slight electron-doping in $d_{z^2}$ orbital, and the similar doping levels in Ni-3$d_{x^2-y^2}$ and interstitial-$s$ orbitals. 
The Ni-3$d_{x^2-y^2}$ and interstitial-$s$ orbitals share the majority of holes. 
However, compared to Fig.~\ref{Fig:orbital_occupancy}(d), neglecting the other orbitals artifically leads to a redistribution of holes between Ni-3$d_{x^2-y^2}$ and interstitial-$s$ orbitals, i.e. some holes transfer from interstitial-$s$ to Ni-3$d_{x^2-y^2}$. 
Downfolding to a 2-band model with only Ni-3$d_{x^2-y^2}$ and interstitial-$s$ further enhances such transfer, resulting in extra holes in Ni-$d_{x^2-y^2}$ orbital  (see Supplementary Information).  
The implication of such hole transfer will be discussed later. 

\begin{figure*}[t]
\centering
\includegraphics[width=\linewidth]{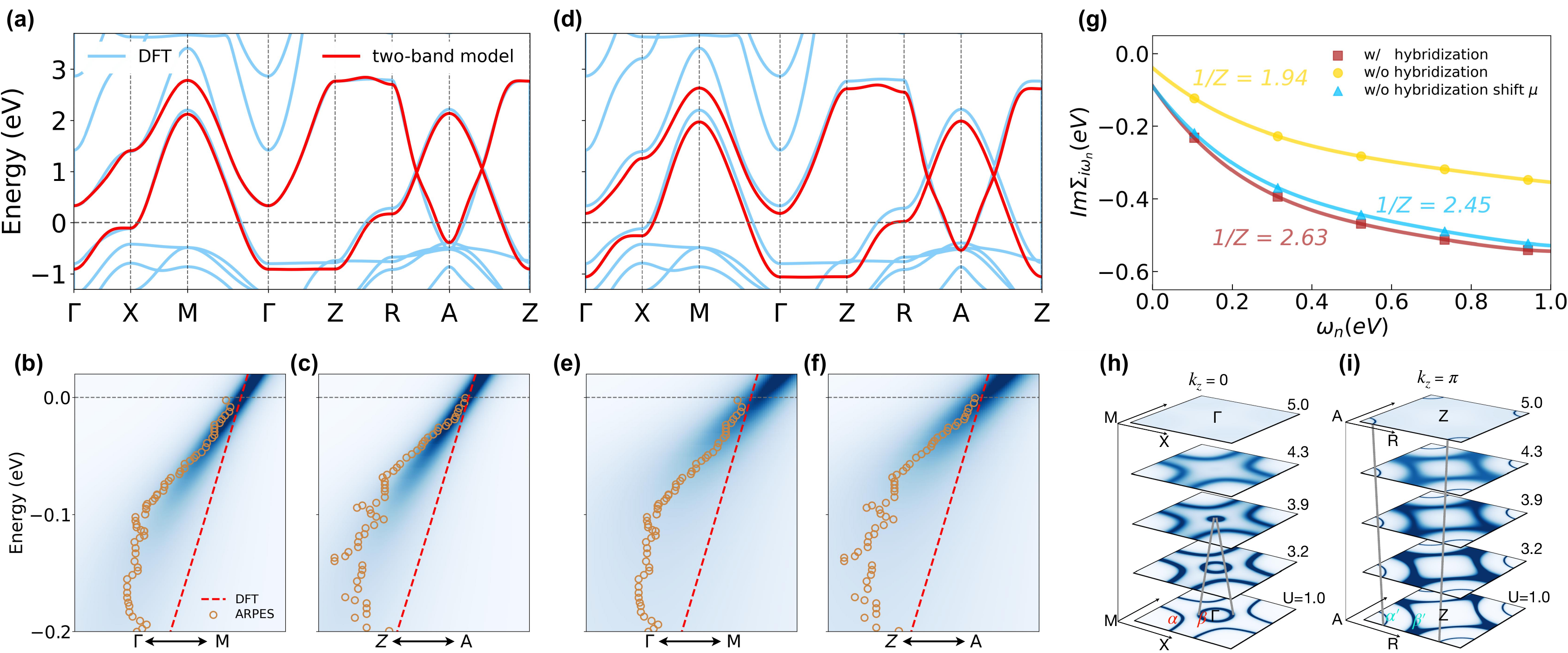}
\caption{\textbf{Role of hybridization between Ni-3$d_{x^2-y^2}$ and interstitial-$s$ orbitals. } (a, d) The comparison of DFT Bloch bands with the 2-band models without Ni-$d_{x^2-y^2}$ and interstitial-$s$ hybridization. Compared to (a), in (b) the chemical potential is shifted up to restore the electron occupancy of 0.75 in Ni-$d_{x^2}-y^2$ orbital. (b, c) The comparison of renormalized electronic structure from DMFT (blue spectra) to DFT (red dashed line) and ARPES (yellow circle) along (b) $\Gamma-M$ and (c) $Z-A$ for model in (a). (e, f) are similar to (b, c) but for the model in (d). (g) The imaginary part of the impurity self-energy of Ni-$d_{x^2-y^2}$ orbital and the corresponding quasiparticle weight. 
(h, i) the Fermi surface evolution of undoped LaNiO$_2$ at (h) $k_z=0$ and (i) $k_z=\pi$ planes, respectively.}
\label{Fig:sd_hybridization}
\end{figure*}

After understanding the hole distributions, we evalute band renormalization in different effective models and compare the results to ARPES. 
In Fig.~\ref{Fig:orbital_occupancy}(f) and (g), the spectra $A(\mathbf{k},\omega)$ colored in blue shows the DMFT electronic structure of the 6 and 2-band model, respectively. 
The red dashed line overlaid on the spectra corresponds to the DFT band structure.  
In both calculations, the Ni-3$d_{x^2-y^2}$ orbital is significantly renormalized by electronic correlation with a clear reduction of band width and single-particle coherence.
In contrast, the interstitial-$s$ orbital remains unaffected by electronic correlation. 
Figure~\ref{Fig:orbital_occupancy}(h) and (i) demonstrate the details of comparison between ARPES and the 2-band model DMFT calculations. 
The DFT band structure has a larger band width and a larger band slope along both $\Gamma-X$ and $Z-R$, while DMFT nicely agrees with ARPES correctly capturing the band renormalization.
The same selective band-renormalization in the 6-band model shown in Fig.~\ref{Fig:orbital_occupancy}(f) indicates that, in addition to the immunity to hole doping, the other Ni-3$d$ orbitals also do not participate in band renormalization. 
For this reason, in the following discussion we will concentrate on the two-band model.   

To understand the influence of interstitial-$s$ orbital on the band renormalization of Ni-3$d_{x^2-y^2}$ orbital, we show in Fig.~\ref{Fig:sd_hybridization}(a) the electronic structure of the 2-band model without hybridization between Ni-3$d_{x^2-y^2}$ and interstitial-$s$ orbitals. 
While the lack of hybridization does not significantly modify the band structure, the renormalization of Ni-3$d_{x^2-y^2}$ orbitals becomes less pronounced. 
As evident in Fig.~\ref{Fig:sd_hybridization}(b) and (c), the slope of the renormalized Ni-3$d_{x^2-y^2}$ band stays between DFT and ARPES, indicating an insufficient renormalization of Ni-3$d_{x^2-y^2}$ orbital.
Evidence from the impurity self-energy and quasiparticle weight shown in Fig.~\ref{Fig:sd_hybridization}(g) further support this conclusion.
Without hybridization, the quasiparticle weight increases from $Z = 1/2.63\approx 0.38$ to $Z = 1/1.94\approx 0.52$. 
This is apparently counter-intuitive to the expectation that the absence of screening from interstitial-$s$ band will lead to a stronger electronic correlation on Ni-3$d_{x^2-y^2}$ band.  
Instead, we found that the unusual enhancement of Ni-3$d_{x^2-y^2}$ correlation stems from the self-doping effect of interstitial-$s$ orbital. 
Turning off the hybridization, electron occupancy in Ni-3$d_{x^2-y^2}$ reduces from 0.75 to 0.65. 
The additional electrons self-doped from interstitial-$s$ to Ni-3$d_{x^2-y^2}$ makes the latter closer to half-filling and, such that a stronger correlation effect is introduced. 
Thus, without hybridization, while Ni-3$d_{x^2-y^2}$ band is less screened, it is more pronouncedly affected by the missing electrons selfly-doped from interstitial-$s$ that results in a reduction of band renormalization. 
The smaller electron occupancy makes  Ni-3$d_{x^2-y^2}$ orbital less correlated. 
To further confirm this, in the non-hybridized 2-band model, we manually shift the chemical potential $\mu$ to restore the electron occupancy 0.75 in Ni-3$d_{x^2-y^2}$ orbital (see Fig.~\ref{Fig:sd_hybridization}(d)) and then performed DMFT calculation. 
The resulting spectra $A(\mathbf{k}, \omega)$ are shown in Fig.~\ref{Fig:sd_hybridization}(e) and (f), which yields a correct band renormalization as compared to ARPES.  
The self-energy and quasiparticle weight shown as cyan triangle line in Fig.~\ref{Fig:sd_hybridization}(g) also closely resemble those of the 2-band model with hybridization (see the brown square line). 
Thus, we conclude that the presence of interstitial-$s$ orbital enhances the electronic correlation of Ni-3$d_{x^2-y^2}$ orbital by self-doping it to near half-filling. 
This also supports the conclusion that the correlated electronic structure of hole-doped LaNiO$_{2}$ can be well described by a single-band model like in cuprates when the correct electron occupancy is maintained~\cite{Kitatani:2020aa, PhysRevX.10.021061, 10.3389/fphy.2021.810394, PhysRevResearch.6.043104, 10.3389/fphy.2022.834682}. 
A recent theoretical study indicates that interstitial-$s$ orbital may mediate superconductivity with the three-dimensional Fermi surface~\cite{xia2025threedimensionalfermisurfacevan}.  

Further evidence on the charge transfer between interstitial-$s$ and Ni-$d_{x^2-y^2}$ orbitals is provided by the Fermi surface evolution driven by electronic correlation, which is supported by the recent experimental observation~\cite{ding_cuprate-like_2024, li2025observationelectridelikesstates}.  
Figure~\ref{Fig:sd_hybridization}(h) and (i) display the Fermi surfaces of the undoped compound LaNiO$_2$ in the $k_z=0$ and $k_z=\pi$ planes under varying correlation strengths $U$. As electronic correlations increase, the $\beta$ electron-pocket near the $\Gamma$ point in the $k_z=0$ plane shrinks and vanishes at $U>3.9$. 
Simultaneously, in the $k_z=\pi$ plane, the $\alpha^\prime$ electron-pocket centered at the A point undergoes a similar contraction with increasing $U$ but remains finite at the experimentally estimated strength $U\sim 4.0$ eV. 
The missing charge in $\beta$ and $\alpha^\prime$ pockets transfer to the Ni-$d_{x^2-y^2}$-derived $\alpha$ and $\beta^\prime$ pockets making them closer to half-filling, which eventually undergo a Mott transition and develop a full charge gap at sufficiently large $U$ (see the Fermi surface for $U=5$ eV.)

\textit{Conclusion} --
In a summary, we studied the hole distribution and the role of interstitial-$s$ orbital in the doped infinite-layer nickelate La$_{1-x}$Sr$_x$NiO$_2$. 
By tracing the distribution of the doped holes, we reveal that they are not doped to Ni-$d_{x^2-y^2}$ and interstitial-$s$ rigidly as expected.
In fact, they are doped nearly with equal amount of holes. 
Based on this crucial observation, we show with DFT + DMFT calculations that, concerning the correlated electronic structure, a single-band model is able to capture the low-energy band renormalization as long as the correct electron occupancy is maintained. 
The presence of interstitial-$s$ orbital shares the doped hole and adjusts the band renormalization of Ni-$d_{x^2-y^2}$ orbital by self-doping effect.

\textit{Acknowledgments} --
This work was supported by the National Key R\&D Program of China (No. 2022YFA1402703, 2023YFA1406400), Sino-German Mobility program (No. M-0006), and Shanghai 2021- Fundamental Research Area (No. 21JC1404700). Part of the calculations was performed at the HPC Platform of ShanghaiTech University Library and Information Services, and the School of Physical Science and Technology.

\bibliographystyle{apsrev4-1}
\bibliography{ref}

\end{document}